\documentstyle[aps]{revtex} 
\begin{document} 
\draft 
\preprint{hep-th/yymmxxx}
\title{`Superresonance' from a rotating acoustic black hole} 
\author{Soumen Basak and Parthasarathi Majumdar} 
\address{The Institute of Mathematical Sciences, Chennai 600 113, 
India\\Email:soumen, partha@imsc.ernet.in} 
\maketitle
\begin{abstract} 
Using the analogy between a shrinking fluid vortex (`draining
bathtub'), modelled as a (2+1) dimensional fluid flow with a sink at the
origin, and a rotating (2+1) dimensional black hole with an ergosphere, it
is shown that a scalar sound wave is reflected from such a
vortex with an {\it amplification} for a specific range of frequencies of
the incident wave, depending on the angular velocity of rotation of the
vortex. We discuss the possibility of observation of this phenomenon,
especially for inviscid fluids like liquid HeII, where vortices with
quantized angular momentum may occur. 
\end{abstract}

The possibility of experimental observation of the acoustic analog of
Hawking radiation from regions of flow of inviscid and barotropic fluids
behaving as {\it outer trapped} surfaces (`acoustic event horizons'), was
pointed out by Unruh \cite{bun} more than two decades ago.  Recent advances
in theoretical understanding of such acoustic analogs of black holes, as
well as improved experimental techniques involving Bose-Einstein condensates
\cite{blv} have brought such experimental possibilities within the realm of
reality. A variety of fluid flow configurations have been identified, all of
which can now be described in terms of different analog black hole models
\cite{vis}.  Many of these can actually be realized in the laboratory under
cryogenic conditions; thus, the analogy enables laboratory tests of a class
of phenomena associated with those aspects of black hole physics that are
insensitive to whether or not the metric satisfies the Einstein equation.

While attention so far has been focussed on the observation of Hawking
radiation from these acoustic black holes, the question of possible energy
extraction from these analogs via the Penrose process \cite{wald} also needs
to be addressed. This process requires black holes with angular momentum. 
The analogous situation would correspond to fluid flow with rotation,
possessing a region of supersonic flow analogous to the {\it ergosphere} of
a Kerr black hole, surrounding the horizon. In the case of the Kerr black
hole, a scalar wave (a solution of the massless Klein-Gordon equation in the
black hole spacetime), entering the ergosphere, is reflected back outside
with an amplitude that exceeds the amplitude of the incident wave, for a
certain range of frequencies bounded from above by the angular velocity of
the black hole.  The wave extracts energy from the rotational energy of the
hole, leading to a `spin-down' of the latter. This wave-analog of the
Penrose process was discovered by Zeldovich \cite{zeld} and investigated
in detail by Starobinsky \cite{star} and Misner (who dubbed the
phenomenon `superradiance') \cite{dew}. 

In this paper, we investigate the possibility of the acoustic analog of
superradiance (a phenomenon that we call `superresonance'), i.e., the
amplification of a sound wave by reflection from the `ergo'-region of a
rotating acoustic black hole. For the latter, we choose the so-called `draining
bathtub' type of fluid flow \cite{vis}, which is basically a 2+1 dimensional
flow with a sink at the origin. A two surface in this flow, on which the
fluid velocity is everywhere pointing towards the sink, and the radial
velocity component exceeds the local sound velocity everywhere, behaves as
an outer trapped surface in this `acoustic' spacetime, and is identified
with the (future) event horizon of the black hole analog. Thus, the velocity
potential for the flow has the form \cite{vis} (in polar coordinates on the
plane)
\begin{equation}
\psi(r,\phi)~=~A ~\log r~+~B~\phi ~, \label{vpot}
\end{equation}
where, $A$ and $B$ are real constants. This leads to the velocity profile
\begin{equation}
{\bf v}={\frac{A}{r}}\hat{r}+{\frac{B}{r}}\hat{\phi} ~. \label{vel}
\end{equation}

It has been shown \cite{vis} that, for barotropic and inviscid fluids with
flows that are free of turbulence, a linear acoustic disturbance is
described by a velocity potential that satisfies the massless
general-coordinate-invariant Klein-Gordon equation,
\begin{equation}
{1 \over \sqrt{-g}} \partial_{\mu}\left( \sqrt{-g}\, g^{\mu \nu}\,
\partial_{\nu} \right) \psi~=~0~, \label{kge}
\end{equation}
where, $g_{\mu \nu}$ is a metric (with Loretizian signature), not of
spacetime itself, but of an acoustic `analog spacetime'. Such analog
spacetimes often resemble black hole spacetimes. For the velocity potential
given in (\ref{vpot}), the analog black hole metric is 2+1 dimensional
with Lorentzian signature, and is given by 
\begin{equation}
{ds}^2=-\left(c^2-\frac{A^2+B^2}{r^2}\right){dt}^2-\frac{2A}{r}\,dr\,dt
-2B\,d\phi\,dt+{dr}^2+r^2\,{d\phi}^2 ~,\label{metr}
\end{equation}
where $c$ is the velocity of sound. 

The properties of this metric  become clearer in slightly different
coordinates, defined through the transformations of the time and
the azimuthal angle coordinates in the {\it exterior} region ($|A|/c +0 \leq r <
\infty$ ),
\begin{eqnarray}
dt ~& \rightarrow &~ dt~+~ {{ |A|~ r} \over {(r^2~c^2-A^2)}}~dr
\nonumber\\
d\phi ~& \rightarrow  &~ d\phi ~+~
{{B~|A|} \over {(r^2~c^2-A^2)}}~dr  ~. ~\label{trans}
\end{eqnarray}
In the new coordinates, the metric in the exterior region assumes the form, after a
rescaling of the `time' coordinate by $c$,
\begin{equation}
{ds}^2=-\left(1-\frac{A^2+B^2}{c^2 r^2}\right){dt}^2 \, + \, 
\left(1-\frac{A^2}{c^2r^2} \right)^{-1} dr^2 \,-\, 2 \, 
\frac{B}{c} \, d\phi \, dt \, + r^2 \, d\phi^2~. \label{met2} 
\end{equation}  

As for the Kerr black hole in general relativity, the radius of the ergosphere is
given by the vanishing of $g_{00}$, i.e., $r_e = (A^2 + B^2)^{1/2}/c$. The metric
has a (coordinate) singularity at $r_h = |A|/c$, which signifies the horizon, i.e.,
the boundary of the outer trapped surface.\footnote{Note that the constant $A$ must
be chosen to be negative to obtain the horizon as indicated.} The metric is a
stationary and axisymmetric one with two Killing vector fields. Note that the
fall-off of the metric components at $r \rightarrow \infty$ is faster than that of
standard asymptotically flat black hole spacetime. This has the consequence that
the standard definitions of globally conserved quantities in terms of Komar
integrals \cite{wald} do not seem to work without modification.

We now turn to the Klein-Gordon equation (\ref{kge}) in the background
metric (\ref{met2}). Given the stationarity and axisymmetry of the metric,
the equation can be separated as 
\begin{equation}
\psi(t, r, \phi)~=~\exp i(\omega t -m \phi) \, R(r)~, ~\label{sol}
\end{equation}
where, $m$ is a real constant that is {\it not} restricted to assume only
a discrete set of values, because we are working with only two space
dimensions. This should be contrasted with the usual 3+1 dimensional
treatment of superradiance \cite{dew}. We assume further that $\omega > 0$. 
The radial function $R(r)$ satisfies the linear second order differential
equation
\begin{eqnarray}
{ 1 \over r} \, \left(1-\frac{A^2}{c^2r^2} \right) \, {d \over {dr}} \left[ r
\, \left(1-\frac{A^2}{c^2r^2} \right) {d \over dr} \right] \, R(r) && 
\nonumber \\
&+& \,\left [\omega^2 - {2Bm\omega \over c r^2} - {m^2 \over r^2} \,
\left(1-{{A^2+B^2} \over {c^2r^2}}\right) \right ]\, R(r) \, = \, 0~.
\label{rad}
\end{eqnarray}
Thus, the problem has reduced to a one dimensional Schr\"odinger problem.  

Two further analytical simplifications can be made: first, we introduce
the tortoise coordinate $ r^{*} $ through the equation
\begin{equation}
\frac{d}{dr^{*}}=\left(1-\frac{A^{2}}{r^{2}~c^{2}}\right)~\frac{d}{dr}  ~,
\label{tor}
\end{equation}  
which implies that 
\begin{equation}
r^{*}=r~+~\frac{|A|}{2~c}~\log\left|\frac{r-\frac{|A|}{c}}{r+\frac{|A|}{c}}\right|~.
\label{tort}
\end{equation}
Observe that the tortoise coordinate spans the entire real line as opposed
to $r$ which spans only the half-line; the horizon $r=|A|/c$ maps to $r^*
\rightarrow -\infty$, while $r \rightarrow \infty$ corresponds to $r^*
\rightarrow +\infty$. Next, introducing a new radial function
$G(r^*) \equiv r^{1/2} \, R(r)$, one obtains the modified radial equation,
\begin{equation}
\frac{d^{2}G(r^{*})}{d{r^{*}}^{2}}~+~\left[Q(r)+\frac{1}{4~r^{2}}~
\left(\frac{dr}{dr^{*}}\right)^{2}
~-~\frac{A^{2}}{r^{4}~c^{2}}~\left(\frac{dr}{dr^{*}}\right)\right]~G(r^{*})=0~,
\label{nrad}
\end{equation}
where,
\begin{equation}
Q(r)=  \frac{A^{2}~m^{2}~+~B^{2}~m^{2}-c^{2}~m^{2}~r^{2}-2~B~m~r^{2}~\omega
~+~r^{4}~\omega^{2}}{r^{4}~}~. \label{que}
\end{equation}  
The main advantage of the new radial equation (\ref{nrad}) over
(\ref{rad}) is the absence in the former of a first derivative of the radial
function; this implies that the Wronskian of linearly independent solutions
is a constant independent of $r^*$ (or $r$). This property is of crucial
importance in what follows. 

We analyze (\ref{nrad}) in two distinct radial regions, viz., near the
horizon, i.e., $r^* \rightarrow -\infty$ and at asymptopia, i.e., $r^*
\rightarrow +\infty$. In the asymptotic region, equation (\ref{nrad}) can be
written approximately as,
\begin{equation}
\frac{d^{2}G(r^{*})}{d{r^{*}}^{2}}~+~\omega^2~G(r^{*})=0~; \label{asym}
\end{equation}
this can be solved trivially,
\begin{equation}
G(r^{*})~=~\exp(i~{\omega}~r^{*})+{\cal R}~
\exp(-i~{\omega}~r^{*})~\equiv G_A(r^*). \label{solr}
\end{equation}
The first term in equation (\ref{solr}) corresponds to ingoing
wave and the second term to the reflected wave, so that ${\cal R}$ is the
reflection coefficient in the sense of potential scattering. It is not
difficult to calculate the Wronskian of the solutions (\ref{solr}); one
obtains
\begin{equation}
{\cal W}(+\infty)~=~-~2~i~\omega~(1-|{\cal R}|^2)~. \label{wro1}
\end{equation}

Near the horizon ($r^* \rightarrow -\infty$),  eqn. (\ref{nrad}) can be
written approximately as,
\begin{equation}
\frac{d^{2}G(r^{*})}{d{r^*}^{2}}~+~\left(\omega 
-m\, \Omega_H \right)^2~G(r^{*})=0 ~. 
\end{equation}
where, $\Omega_H \equiv Bc/A^2$ is identified with the angular velocity
of the acoustic black hole. We impose the boundary condition that of
the two solutions of this equation, only the {\it ingoing} one is physical,
so that one has
\begin{equation}
G(r^*)~=~{\cal T}~\exp i(\omega - m \Omega_H)~r^*~\equiv ~G_H(r^*) ~.
\label{hor}
\end{equation}
The undetermined coefficient ${\cal T}$ is the transmission coefficient of
the one dimensional Schr\"odinger problem. Once again, it is easy to
calculate the Wronskian of this solution; one obtains
\begin{equation}
{\cal W}(-\infty)~=~-~2~i~(\omega~-~m\Omega_H)~|{\cal T}|^2~.
\label{wro2}
\end{equation}
Since both equations are actually limiting approximations of
eq. (\ref{nrad}), which, as we have mentioned, has a constant Wronskian,
it follows that
\begin{equation}
{\cal W}(+\infty)~=~{\cal W}(-\infty)~, \label{wroe}
\end{equation}
so that, from eqn.s (\ref{wro1}) and (\ref{wro2}), we obtain the relation
\begin{equation}
1~-~|{\cal R}|^2~=~\left(~{{\omega - m \Omega_H} \over \omega}~\right)~|{\cal
T}|^2~. \label{main}
\end{equation}
It is obvious from eq. (\ref{main}) that, for frequencies in the range $0
< \omega < m \Omega_H$, the reflection coefficient has a magnitude {\it
larger than unity}. This is precisely the amplification relation that
emerges in superradiance from rotating black holes in general relativity
\cite{star}, \cite{dew}. 

For a given frequency $\omega$ in the above range and a given value of
the azimuthal mode number $m$, the total energy flux at asymptopia can be
easily calculated \cite{bas} in terms of the reflection coefficient
${\cal R}$. By conservation of energy, this flux must equal the rate of
loss of mass (energy) from the black hole, so that one may write
\begin{eqnarray}
{dM \over dt}&~=~&-~{\cal P}^{\infty}_{\omega m}~=~{\pi \omega^2 \over
c}~\left(~1~-~|{\cal R}_{\omega m}|^2~\right)~\nonumber \\
&~=~&{\pi \omega \over
c}~\left(\omega~-~m~\Omega_H \right)~|{\cal T}_{\omega m}|^2~,
\label{flux}
\end{eqnarray}
where, $M$ is the mass of the acoustic black hole (which must be related
to the energy of the vortex under consideration) and ${\cal
P}^{\infty}_{\omega m}$ is the average power radiated to asymptopia for a
given $\omega$ and $m$ (modulo an overall positive constant proportional
to the equilibrium density of the fluid). We have used eq. (\ref{main}) in
arriving
at (\ref{flux}). The loss of mass of the black hole superresonance is
obvious in the above equation, so long as the frequency lies in the
range quoted above. A similar equation can be derived for the time rate
of change of angular momentum of the hole \cite{dew}. 

In order to link up more precisely with experimental possibilities, one
needs to be able to compute the reflection coefficient by explicitly
solving eq. (\ref{nrad}). We shall not attempt this excercise in this
paper, but postpone that discussion to a future publication \cite{bas}. 
Here, we shall concentrate instead on certain qualitative features of the
sort of observations one might hope to make on this phenomenon. In
particular, our assumption at the outset of an irrotational, nonviscous
fluid flow leads us immediately to a kind of fluid that is known for
decades to have truly remarkable properties, liquid HeII, or more
precisely, superfluid He4. We are especially interested in the existence
of vortices with quantized angular momenta. We follow \cite{fey} to
obtain a rough picture of what this will do to superresonant
amplitudes.

The velocity profile for fluids under consideration is given by eq.
(\ref{vel}) and follows from (\ref{vpot}) using standard relations of
fluid dynamics. Now, let us imagine that the fluid is actually superfluid
HeII and our black hole is a vortex in the fluid with a sink at the
centre. In the quantum theory of HeII, the wave function is of the form
\begin{equation} 
\Psi ~=~\exp i~ \sum_i~\phi({\vec r}_i)~\Phi_{ground}~ , \label{wav}
\end{equation}
where, ${\vec r}_i$ is the location of the $i$th particle of HeII. The
velocity at any point is given by the gradient of the phase at that point
\begin{equation}
{\bf v}~\equiv~{\bf \nabla}~\phi~. \label{vel2}
\end{equation}
This means that the velocity potential can be thought of crudely as the
phase of the wave function. However, this phase cannot be non-singular
everywhere, since we are assuming a sink at $r=0$. Thus, there exist
non-trivial holonomies of the velocity field on closed curves around this
sink. For simplicity, let us concentrate on circles around the sink. The
change of phase of the wave function around any of these circles, for the
velocity profile given in (\ref{vel}) is given by
\begin{equation}
\Delta \phi ~\propto~2\pi~B~, \label{del}
\end{equation}
where, the constant of proportionality depends on the microscopic
properties of the fluid, and is unaffected by our choice of (\ref{vel}). 
Thus, for the overall wave function to be single valued, the parameter
$B$ in the acoustic black hole metric must be proportional to an integer
$n$. 

The immediate implication of this property is that the angular velocity
$\Omega_H$ at the horizon is likewise restricted to be proportional to the
integer $n$. The likely fallout of this is `quantization' of the outgoing
energy flux, through the last part of eq. (\ref{flux}). In other words,
it is conceivable that the angular momentum of the black hole will now
change discontinuously from being proportional to $n$ to being
proportional to $n-1$ due to superresonance; this will cause the energy
flux also to change in discrete steps - an effect that should make
observations easier than in actual black holes. This will also be
manifest in the frequency spectrum of the reflection coefficient ${\cal
R}$ \cite{alp} : one should see a multilpicity of equally-spaced peaks in
this spectrum, perhaps with differing strengths and, of course, with
widths that are multiples of the smallest width. We hope to report in
more quantitative terms on these interesting aspects of superresonance in
a forthcoming publication. 

We end the letter with the rather curious observation that at the time of
the original discovery of superradiance, the phenomenon of sound
amplification due to reflection from a medium at rest, with a
supersonically moving boundary, was already known for about four decades
\cite{zeld}.  It is not completely clear to us whether this is the same
as superresonance.  There may also be some connection of superresonance
with stimulated vortex sound \cite{morin}. If indeed these classical
acoustic phenomena are examples of superresonance, then one would be led
to conclude that evidence already exists of acoustic black hole analogs
existing in nature. On the other hand, it also means that these acoustic
phenomena have a novel interpretation in terms of semiclassical black
hole analog models. However, the intriguing aspects of superresonance in
HeII may still be novel enough to do further experiments, to confirm all
aspects of the black hole analog picture. 

This is a preliminary report of an ongoing investigation that began in
collaboration with S. Das and G. Kunstatter; One of us (P.M.) thanks them
both for stimulating initial discussions and also for pointing out
\cite{alp} to us. He also thanks the Physics Department of the University
of Winnipeg for hospitality during the initial phase of this work. We
thank R. Bhaduri and G. Date for very illuminating discussions.

\noindent{\it Note added} After submission of this paper to the Archives,
we became aware of some general discussion of the possibility of
superresonance in superfluid HeII \cite{vol}; however, the aspects of the
phenomenon dealt with in that reference seem to be different from our
concerns in this paper. We thank G. Volovik for bringing this reference to our
attention. We also thank U. Fischer for correspondence regarding transformations
(\ref{trans}) and for bringing \cite{fis} to our attention. Reservations
expressed about the validity of these tranformations do not appear to be
significant in the region {\it exterior} to the event horizon, which is our region
of interest.

\end{document}